\documentclass[aps,prl,floatfix,epsfig,twocolumn,showpacs,preprintnumbers]{revtex4}
\usepackage{amssymb}
\usepackage{graphicx}
\usepackage{amsmath}

\input{tcilatex}
\begin{document}

\title{Calculated Magnetic and Electronic Properties of Pyrochlore Iridates}
\author{Xiangang Wan$^{1}$, Jinming Dong$^{1}$, Sergey Y. Savrasov$^{2}$}
\affiliation{$^{1}$National Laboratory of Solid State Microstructures and Department of
Physics, Nanjing University, Nanjing 210093, China\\
$^{2}$Department of Physics, University of California, Davis, One Shields
Avenue, Davis, Ca 95616}
\date{\today}

\begin{abstract}
Using density functional theory and LDA+U method, we investigate magnetic
and electronic structure of Y$_{2}$Ir$_{2}$O$_{7}$ and rare--earth based
pyrochlore iridates. Our study reveals that the ground state is a
non--collinear magnetic insulating state. Due to strong spin-orbit coupling
in Ir 5\textit{d}, there is an unusual correlation between the bands near
Fermi level and the magnetization direction, resulting in a possibility of
insulator--to--metal transition under applied magnetic field. This makes\
pyrochlore iridates a good candidate for possible magnetoressitance and
magnetooptical applications.
\end{abstract}

\date{\today }
\maketitle

\input{epsf}

While it has been realized that the Coulomb interaction is of substantial
importance in 3\textit{d} transition metal systems, resulting in a great
variety of their physical properties, such, e.g., as metal--insulator
transition\cite{Review M-I}, colossal magnetoresistance\cite{CMR} and high--T%
$_{c}$ superconductivity\cite{Pickett-HTC}, the 4\textit{d} and the 5\textit{%
d} orbitals are, on the other hand, spatially more extended and usually
possess a much broader bandwidth, making the correlation effects to be
minimal. However, very recently, both theory and experiment give the
evidence on the importance of interplay between spin--orbit coupling and
Coulomb interaction for several, mainly Ir based, 5\textit{d }compounds,
which have received a lot of research attention \cite%
{Sr2IrO4-1,Sr2IrO4-2,Y2Ir2O7 U,GME,Iridate-2,Iridate 3,Iridate,Pickett,Os
compound,Spin-Hall Na2IrO3 e-e and SO, Leon e-e and SO}. In particular, in Sr%
$_{2}$IrO$_{4}$, one of most studied Ir oxides, it has been found that the
material becomes a Mott insulator \cite{Sr2IrO4-1,Sr2IrO4-2}, and exhibits
very interesting magnetoelectric properties\cite{GME}. A different system, Na%
$_{4}$Ir$_{3}$O$_{8}$, which crystallizes in a geometrically frustrated
spinel structure, is an insulator with a large Curie--Wiess temperature
(about 650 K) and with a considerable effective moment (1.96 $\mu _{B}$)\cite%
{Spin-liquid Na4I3O8 -exp}. However, this compound does not exhibit any sign
of magnetic order even for the lowest measured temperature\cite{Spin-liquid
Na4I3O8 -exp}, and it has been suggested as one of the few long sought
quantum spin liquids\cite{Spin-liquid,S-L-3,S-L -2}. There is also
theoretical work that addresses anisotropy of magnetic interactions\cite%
{A-J,Anisotropy J}.

Ir oxides \textit{A}$_{2}$Ir$_{2}$O$_{7}$\cite{Y2Ir2O7 U,Matsuhira
Ir-227,exp 2001 Ir-227,Ir-227,Taira 2001 Ir-227,Kondo in Pr2Ir2O7,structure
R2Ir2O7} (\textit{A}= Y or rare--earth element), which crystallize in a
pyrochlore structure\cite{pyrochlore review}, is another geometrically
frustrated iridate system. It has been recently discussed in connection with
a novel "topological Mott insulator" phase\cite{Leon e-e and SO} seen
between topological band insulating\cite{TBI} and conventional Mott
insulating\cite{Review M-I} phases as interactions get stronger. Experiment
observes that depending on the \textit{A--}site, \textit{A}$_{2}$Ir$_{2}$O$%
_{7}$ show a wide range of electrical properties \cite{Matsuhira Ir-227,exp
2001 Ir-227,Ir-227,Taira 2001 Ir-227,Kondo in Pr2Ir2O7}. For example, Y$_{2}$%
Ir$_{2}$O$_{7}$ is an insulator\cite{Y2Ir2O7 U} but with increasing the
ionic radius at the \textit{A}--site, the system eventually becomes metallic
for Nd$_{2}$Ir$_{2}$O$_{7}$\cite{Matsuhira Ir-227}, while Pr$_{2}$Ir$_{2}$O$%
_{7}$ shows strong Kondo behavior\cite{Kondo in Pr2Ir2O7}. Moreover, it has
been found that temperature will drive an insulator--to--metal transition
associated with abnormal magnetic behavior without structural change\cite%
{Matsuhira Ir-227}. There are also several electronic structure calculations
trying to explore electronic and magnetic properties of the iridates with
geometrically frustrated structure\cite{A-J, LSDA+U Y2Ir2O7}.

In the present work, we perform a detailed study of the magnetic and
electronic structure for \textit{A}$_{2}$Ir$_{2}$O$_{7}$. We find that the
ground state of those systems is a non--collinear magnetic state while
geometrically frustrated pyrochlore lattice makes other states with
different orientations of moments close in energy. When \textit{A}=Pr and Nd
the electronic structure of these compounds shows metallic behavior while
for \textit{A}=Y, Sm, Eu the materials are insulators. We uncover that
spin--orbit (SO) coupling affects the energy bands near the Fermi level and
depending on orientation of moments some of those systems can be switched
from an insulator to a metal. Exotic electronic and magnetic properties
would make Ir pyrochlore to be good candidates for various applications
including magnetoresistance effect and magnetooptics.

We perform our electronic structure calculations based on local spin density
approximation (LSDA) to density functional theory (DFT) with the
full--potential, all--electron, linear--muffin--tin--orbital (LMTO) method%
\cite{FP-LMTO}. We use LSDA+U scheme\cite{LDA+U} to take into account the
electron--electron interaction between Ir 5\textit{d }electrons, and use U =
2 eV which has been previously found to be adequate in iridates\cite%
{Sr2IrO4-1,Sr2IrO4-2,Iridate 3}. When the \textit{A} site is a rare earth
element, we also add the Coulomb interaction for the localized 4\textit{f}
electrons and use U = 6 eV. We use a 12$\times $12$\times $12 k--mesh to
perform Brillouin zone integration, and switch off symmetry operations in
order to minimize possible numerical errors in studies of various
(non--)collinear configurations. We use experimental lattice parameters\cite%
{Taira 2001 Ir-227} in all set ups.

We first discuss our results for Y$_{2}$Ir$_{2}$O$_{7}$. Without spin orbit
(SO) coupling, both LSDA and LSDA+U predict this system to be metallic which
is not consistent with the experiment. Since the SO strength is large for Ir
5\textit{d} electrons (about 0.4 eV)\cite{SO constant}, and has been found
to produce an insulating behavior in Sr$_{2}$IrO$_{4}$\cite%
{Sr2IrO4-1,Sr2IrO4-2}, we have performed the LSDA+U+SO calculations. There
are four Ir atoms inside the unit cell forming a tetrahedral network as
shown in Fig.1. We first set an initial magnetization axis along (001)
direction, however the calculations converge to a non--collinear state with
the magnetic moment departing from the initial orientation a little bit. The
four Ir sites have similar spin and orbital moments (about 0.13 $\mu _{B}$)
and both of them are slightly\ smaller than theoretical values reported
previously\cite{LSDA+U Y2Ir2O7}. However, due to the obtained slight
non--collinearity of the solution, the net magnetic moment is found to be
small.

Each of four Ir atoms is octahedrally coordinated by six O\ atoms, which
makes the Ir 5\textit{d} state split into doubly degenerate e$_{g}$ and
triply degenerate t$_{2g}$ states. Due to the extended nature of Ir 5\textit{%
d} orbital, the crystal--field splitting between t$_{2g}$ and e$_{g}$ is
large with the e$_{g}$ band to be 2 eV higher than the Fermi level. The
bands near the Fermi level are mainly contributed by Ir t$_{2g}$ with some
mixing with O 2\textit{p} states. SO coupling has a considerable effect on
these t$_{2g}$ states: it lifts their degeneracy and produces 24 separate
bands in the range from -2.3 to 0.7 eV. Same as in Sr$_{2}$IrO$_{4}$\cite%
{Sr2IrO4-1,Sr2IrO4-2}, the bandwidth of these t$_{2g}$ states in our
LSDA+U+SO calculation for Y$_{2}$Ir$_{2}$O$_{7}$ is also narrow, however, it
is still metallic as shown in Fig.2(a). Naively one may expect that using
larger Coulomb U\ will result in an insulating state. However, our
additional calculations show that increasing U\ cannot solve this problem,
and even a quite large U (=5 eV) cannot open a band gap for the initial
collinear (001) setup. Here we agree with the previous calculation\cite%
{LSDA+U Y2Ir2O7}, which also showed that U\ cannot open a band gap.

Strong spin--orbit coupling may however induce the dependence of the band
structure on quantization axis. So, we subsequently perform our calculations
with initial magnetization aligning along (110), (120), (111) directions. We
also perform the calculations with two sites in a tetrahedron along and
other two pointed oppositely to (001), (111), (110) or (120) direction in
order to account for possible antiferromagnetism. Interestingly, we find
that despite the strong SO coupling which usually results in a large
magnetic anisotropy energy, for Y$_{2}$Ir$_{2}$O$_{7}$, the rotation of
magnetization does not involve too much change in the total energy. Among
the magnetic configurations mentioned above, the (111) direction is found to
be lowest, but the energy difference between this and the highest energy
(001)\ state is just about 3.7 meV per unit cell. It is also interesting
that while the magnetization direction does have a considerable effect on
the bands near the Fermi level, as can be seen from comparing the Fig.2(a)
and Fig.2(b), it turns out that all of the above mentioned calculations
converge to metallic states. Also, all of them produce a considerable net
magnetic moment in contrast to the experiment \cite{exp 2001 Ir-227, Taira
2001 Ir-227, Ir-227}. Thus, one can conclude that neither of these magnetic
configurations may be the ground state. 
\begin{table}[tbp]
\caption{The spin $\langle S\rangle $ and orbital $\langle O\rangle $\
moment (in $\protect\mu _{B}$), the total energy E$_{tot}$ (in meV) as well
as the band gap E$_{gap}$(in meV) for several selected magnetic
configurations of Y$_{2}$Ir$_{2}$O$_{7}$. }%
\begin{tabular}{ccccc}
\hline
Configuration: & (001) & (111) & 2--in/2--out & all--in/out \\ \hline
$\langle $S$\rangle $ & 0.13 & 0.17 & 0.14 & 0.13 \\ 
$\langle $O$\rangle $ & 0.13 & 0.13 & 0.15 & 0.17 \\ 
E$_{tot}$(meV) & 4.47 & 0.69 & 3.21 & 0.00 \\ 
E$_{gap}$(meV) & 0 & 0 & 0 & 30 \\ \hline
\end{tabular}%
\end{table}

With the pyrochlore structure, the Ir sublattice has a topology consisting
of corner--sharing tetrahedra and is geometrically frustrated. Thus, we
carry out several non--collinear calculations with the initial state to be
"all--in/out" (where all moments point to or away from the centers of the
tetrahedron), "2--in/2--out" (two moments in a tetrahedron point to the
center of this tetrahedron, while the other two moments point away from the
center, i.e. the spin--ice\cite{Spin ice Ln2Ti2O7} configuration), and
"3--in/1--out" magnetic structures (see Fig.1 for the moments
configuration). We find that the "all--in/out" configuration is the ground
state, and this state is insulating as shown in Fig.2(c): it has a band gap
of 0.03 eV. The experimental band gap for Y$_{2}$Ir$_{2}$O$_{7}$ is not
available, though it had been expected that it should be larger than that of
Sm$_{2}$Ir$_{2}$O$_{7}\ $(0.01 eV)\cite{Matsuhira Ir-227}. Different from
other magnetic configurations, during the self--consistency the
"all--in/out" state will retain their initial input direction; thus, there
is no net magnetic moment. This is consistent with the experimental fact on
the absence of the magnetic hysteresis loop\cite{Taira 2001 Ir-227}.\ As
shown in Table I, the energy difference between the ground and several
selected excited states with different orientations of moments is small.
Here, the energy difference involves not only the energy of magnetic
anisotropy but also the effect of intersite exchange interaction. Due to the
geometrically frustrated structure this energy difference is obviously
smaller than that found for another 5\textit{d} compound Ba$_{2}$NaOsO$_{6}$%
\cite{Os compound}.

Regardless the proximity of the ground and excited (or said metastable)
states in energy, they would have very different conductivity and magnetic
properties. Thus one can understand that for the same compound Y$_{2}$Ir$%
_{2} $O$_{7}$ Taira \textit{et al.}\cite{Taira 2001 Ir-227} observe no
ferromagnetic ordering while Yanagishima \textit{et al.}\cite{exp 2001
Ir-227,Ir-227} claim a presence of a small net magnetic moment. One can also
understand the observed temperature induced metal--insulator transition, as
well as a large difference in temperature dependence of magnetization
measured under zero--field--cooled conditions (ZFC) and under field--cooled
conditions (FC) at low T's.

\begin{figure}[tbp]
\includegraphics [height=1.8in] {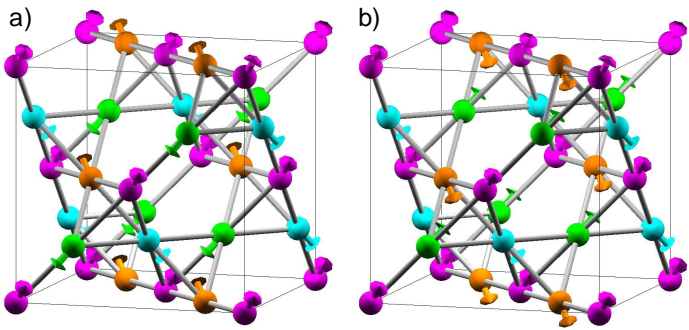}
\caption{ The pyrochlore structure showing Ir tetrahedral network and its
magnetic configurations (a) "all--in/out" configuration. (b) "2--in/2--out"
configuration. The arrows denote moments directions.}
\end{figure}

Based on both strong sensitivity of the energy bands near the Fermi level on
the orientation of moments, and proximity of various magnetic states in
energy, it is natural to expect that an application of a magnetic field
could have a big effect not only on the magnetic response but also on the
conductivity in iridates. In particular, this should result in a\ large
magnetoresistance effect if one is able to switch between insulating
"all--in/out" state and any collinear state. This simple idea has been
proved by the following numerical calculation. Starting from the
"all--in/out" ground state, we apply an external field along (001)
direction. The result shows that the external field will rotate the magnetic
moments meanwhile only slightly change their magnitude. A 5 T magnetic field
along (001) induces a 0.07 $\mu _{B}$\ net magnetic moment, which is in fact
close to the experiment performed for Sm$_{2}$Ir$_{2}$O$_{7}$, where it was
shown that a 4 T magnetic field produces a 0.05 $\mu _{B}$\ total moment\cite%
{Taira 2001 Ir-227}. Increasing the field further, the numerical calculation
does find energy bands crossing the Fermi level, namely an
insulator--to--metal transition at a field of 40 T, although there is
already non--negligible density of states at E$_{f}$ for a lower magnetic
field.

\begin{figure}[tbp]
\includegraphics [height=4.5in] {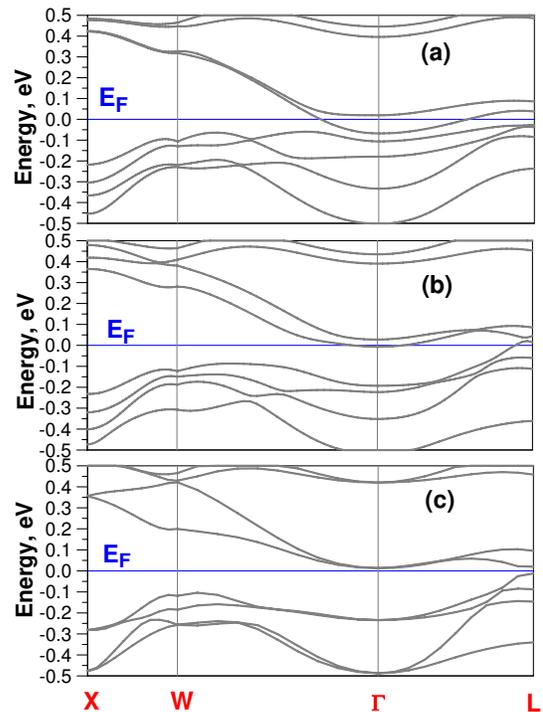}
\caption{LSDA+U+SO electronic band structure of Y$_{2}$Ir$_{2}$O$_{7}$ for
different orientations of magnetic moments: (a) along (001) direction; (b)
along (111) magnetization direction; (c)\ "all-in/out" configuration.}
\end{figure}

Another interesting feature which emerges from this electronic structure is
an expected magnetic field dependent optical response. We have computed
interband optical conductivity $\sigma _{inter}(\omega )$ of Y$_{2}$Ir$_{2}$O%
$_{7}$ for (001), (111) and "all--in/out" orientations of moments. The
results are shown on Fig.3 where one can see a very different behavior of
this function depending on the imposed magnetic configuration at infrared
frequencies. The intraband Drude contribution can also be restored from our
computed intraband plasma frequencies for the metallic (001) and (111)
states which are pretty small and equal to 0.7 and 0.2 eV, respectively.
Thus, controlling the strength and direction of the applied magnetic field
one can reach a continuous change from metallic to insulating response in
this system.

\begin{figure}[tbp]
\includegraphics [height=2.5in] {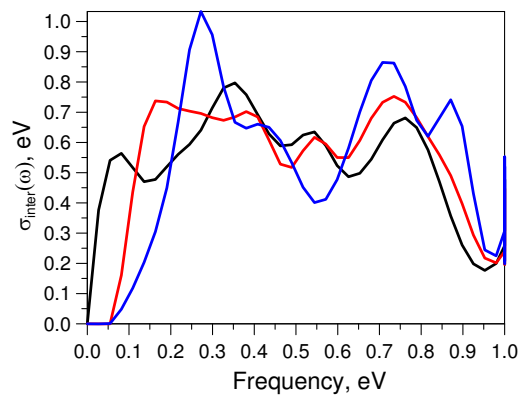}
\caption{Calculated interband optical conductivity of Y$_{2}$Ir$_{2}$O$_{7}$
for different orientations of magnetic moments: black -- along (001)
direction; red -- along (111) magnetization direction; blue --\ "all-in/out"
configuration.}
\end{figure}

Experiment finds that changing the \textit{A}--site of \textit{A}$_{2}$Ir$%
_{2}$O$_{7}$ will vary the properties considerably. Yanagishima and Maeno 
\cite{exp 2001 Ir-227} show that for \textit{A}$_{2}$Ir$_{2}$O$_{7}$, the
electrical conductivities of \textit{A}=Pr, Nd, Sm and Eu exhibit metallic
behavior, while the electrical conductivities of \textit{A}=Gd, Tb, Dy, Ho,
Yb and Y are like in insulators. Matsuhira \textit{et al.}\cite{Matsuhira
Ir-227} also observe that the \textit{A}--site induces the change from metal
to insulator, but they claim that it is taken place around Nd. Namely, the
systems with \textit{A}=Nd, Sm and Eu are insulators while those with 
\textit{A}=Pr are metals. Matsuhira \textit{et al.}\cite{Matsuhira Ir-227}
try to contribute this discrepancy to the quality of samples. However, even
using the same synthesis condition, the discrepancy still appears\cite%
{Matsuhira Ir-227}.

To clarify the effect of \textit{A}--site, we first perform a constrained
calculation with the 4\textit{f} band shifted by a constrained potential.
This however almost does not affect the bands around the Fermi level. So,
one can conclude that the rare earth element has a small effect on the
conductivity. We further perform the calculation by using \textit{A}$_{2}$Ir$%
_{2}$O$_{7}$ structure but replacing the rare earth element \textit{A} by Y.
Like in Y$_{2}$Ir$_{2}$O$_{7}$, the ground state is found to be
"all--in/out" non--collinear solution. However for Pr$_{2}$Ir$_{2}$O$_{7}$
and Nd$_{2}$Ir$_{2}$O$_{7}$, the ground state is metallic. Changing the 
\textit{A}--site from Nd to Sm and further to Eu results in decrease in the
ionic radius. This reduces the Ir--O--Ir angle, and makes our calculation
for \textit{A}=Sm and Eu to produce insulating band structures. We
contribute the discrepancy between Ref.\cite{exp 2001 Ir-227} and Ref.\cite%
{Matsuhira Ir-227} to the smallness of the energy difference between the
"all--in/out" ground state and other metastable states.

In summary, using the LSDA+U+SO method, we have explored the electronic and
magnetic properties of geometrically frustrated pyrochlore iridates. Our
study reveals that the ground state of these systems has a non--collinear
"all--in/out" magnetic configuration of moments. Our constrained calculation
shows that the \textit{A} site has a small affect on the energy bands near
the Fermi level while different ionic radii and corresponding changes in
lattice parameters are the reason why \textit{A}=Pr are metallic while 
\textit{A}=Nd, Sm, Eu, and Y are insulating. Thanks to the strong SO
coupling in Ir 5\textit{d }states, rotation of magnetization can produce
insulator--to--metal transition thus making iridates to be interesting
candidates for various applications, such, \textit{e.g.}, as giant
magnetoresistance effect and magnetooptics.

The work was supported by National Key Project for Basic Research of China
(Grant No. 2006CB921802, and 2010CB923404), NSFC under Grant No. 10774067,
and 10974082. We also acknowledge DOE SciDAC Grant No. SE--FC02--06ER25793
and KITP where part of the work has been performed.

\end{document}